\documentclass[fleqn,twoside]{article}
\usepackage{espcrc2}

\usepackage{graphicx}
\usepackage[figuresright]{rotating}

\newcommand{\AmS}{{\protect\the\textfont2
  A\kern-.1667em\lower.5ex\hbox{M}\kern-.125emS}}

\title{Archeops, mapping the CMB sky from large to small angular scales}

\author{J.-Ch. Hamilton
\address{Institut des Sciences Nucl{\'e}aires, CNRS-IN2P3, Grenoble, France}
\address{Physique Corpuculaire et Cosmologie, CNRS-IN2P3, Coll{\`e}ge de France,
  Paris, France}, on behalf of the Archeops Collaboration.
}
       
\begin{document}

\begin{abstract}
  Archeops is a balloon-borne experiment designed to measure the
  temperature fluctuations of the CMB on a large region of the sky
  ($\simeq 30\%$) with a high angular resolution (10 arcminutes) and a
  high sensitivity ($60\mu\mathrm{K}$ per pixel). Archeops will perform a
  measurement of the CMB anisotropies power spectrum from large
  angular scales ($\ell\simeq 30$) to small angular scales ($\ell
  \simeq 800$). Archeops flew for the first time for a test flight in
  July 1999 from Sicily to Spain and the first scientific flight took
  place from Sweden to Russia in January 2001. The data analysis is on
  its way and I present here preliminary results, realistic
  simulations showing the expected accuracy on the measurement of the
  power spectrum and perspectives for the incoming flights (Winter
  2001/2003).
 \vspace{1pc}
\end{abstract}

\maketitle

\section{Introduction}
\subsection{CMB physics}
In the framework of Big-Bang theory, the Universe started with a hot
and dense phase about 15 billion years ago and cooled down while
expanding. The first neutral atoms formed when the temperature was
about 13.6 eV (160000 K), but due to the large number of photons
compared to baryons (ratio $\simeq 10^9$), the Universe remained
ionized until the temperature dropped below 0.3 eV (3000 K). At this
moment, the mean free path of the photons increased drastically so
that the photons that scattered at this time have not interacted with
matter since then. This moment is known as {\em matter-radiation
  decoupling} or {\em recombination}. Those photons cooled down with
the expansion of the Universe and are know observed at a temperature
of 2.7 K. As the matter and radiation were at thermal equilibrium
before decoupling, these photons have a pure blackbody spectrum and
are homogeneously distributed on the celestial sphere. This radiation
is known as the {\em Cosmic Microwave Background} (hereafter CMB).

The discovery of the CMB by Penzias and Wilson~\cite{penzias_wilson}
and its interpretation in terms of a Big-Bang relic by Dicke and
collaborators~\cite{dicke} was a major argument for the Big-Bang
theory~\cite{gamow,alpher_herman}. The CMB temperature was measured to
be highly isotropic but tiny anisotropies were expected. These
temperature fluctuations reflect the density fluctuations on the last
scattering surface.  These are necessary to explain the presence of
structures in the Universe such as galaxies and clusters. The CMB
anisotropies were discovered by the COBE satellite with a {\em rms}
amplitude of about 30 $\mu\mathrm{K}$~\cite{smoot} at scales larger
than 7 degrees. COBE also measured its spectrum with high
precision~\cite{mather,fixsen} proving its pure blackbody nature.

The CMB anisotropy typical physical size in the last scattering
surface can be theoretically predicted while its angular size as seen
from here and now depends on the geometry of the Universe along the
path of the photons. Hence, mapping the CMB anisotropies is a powerful
cosmological test.

The two competing paradigms for the origin of structures in the
Universe, namely inflation and topological defects, predict
significantly different distributions for the former density
fluctuations. These distributions propagate to us in a cosmological
parameters dependent way to describe the temperature anisotropies that
we expect on the sky\footnote{Freely available numerical codes, such
  as CMBFast~\cite{cmbfast}, have been developed for this purpose.}.
It is therefore of deep interest to investigate their angular
distribution and compare the measurements to cosmological models.

\begin{figure}[htb]
\resizebox{\hsize}{!}{\includegraphics{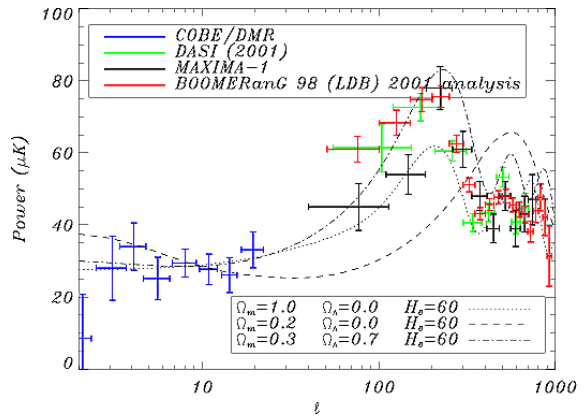}}
\caption{Expected CMB power spectrum 
  $\left(\sqrt{\frac{\ell\left(\ell+1\right)C_\ell}{2\pi}}\right)$ for
  inflationary-like primordial density fluctuations (black curves) for
  three different cosmological models along with the latest
  measurements from BOOMERanG, MAXIMA and DASI and the earlier
  measurements from COBE.}
\label{speccmb}
\end{figure}

The temperature anisotropies on the sky are commonly described via their spherical harmonics expansion,
\begin{equation}
\frac{\delta T}{T}\left(\theta,\phi\right)=\sum_{\ell=0}^\infty\sum_{m=-\ell}^\ell a_{\ell m}Y_{\ell m}\left(\theta,\phi\right),
\end{equation}
where $\ell$ is the multipole index, inversely proportional to the
angular scale (1 degree roughly corresponds to $\ell=200$). The angular power spectrum of the temperature fluctuations of the CMB is defined as:
\begin{equation}
C_\ell=\frac{1}{2\ell+1}\sum_{m=-\ell}^\ell\left|a_{\ell m}\right|^2.
\end{equation}
The evolution of the angular power spectrum of the CMB as a function of $\ell$ can be splitted into three major regions (see Figure~\ref{speccmb}):
\begin{itemize}
\item On the low-$\ell$ part (large angular scales) no particular
  structure is expected as we are considering physical sizes on the
  last scattering surface larger than the horizon at the epoch of
  decoupling. No physical process is expected to have modified those
  fluctuations since the early Universe.
\item Between $\ell\simeq 30$ and $\ell\simeq 1000$ (degree and
  sub-degree scales) we a re considering structures that had time to
  collapse and experience acoustic oscillations between the
  matter-radiation equality and the matter radiation decoupling. We
  therefore expect a series of acoustic peaks (the first one being
  located around $\ell=200$, corresponding to the size of the horizon
  at the epoch of decoupling) in the case of inflationary-like early
  Universe models where the oscillations are in phase. In the case of
  isocurvature fluctuations (such as topological defects), the
  oscillations are not in phase and a large bump is expected, but no
  multiple peaks.
\item In the large $\ell$ part (arcminute scales and below), the power
  is expected to drop drastically due to the finite thickness of the
  last scattering surface and to the finite value of the mean free
  path of the photons before decoupling.
\end{itemize}

\subsection{Recent results}
Our knowledge of the CMB power spectrum has been significantly
improved since COBE measurements~\cite{smoot} by two balloon-borne
experiments, BOOMERanG~\cite{netterfield} and MAXIMA~\cite{hanany},
and a ground based interferometric experiment, DASI~\cite{halverson}.
These results are shown in Figure~\ref{speccmb}. We observe that the
very low part of the power spectrum is highly constrained by the COBE
points while the high-$\ell$ acoustic region is constrained by the
three recent experiments showing undoubtly the multiple peak feature
that is expected from inflation. This set of data therefore strongly
disfavors topological defects as seeds for the structure formation in
the Universe. Comparing this set of data with theoretical power
spectra have lead to the estimation of the cosmological
parameters~\cite{netterfield,balbi,bond}. The favored model is
dominated by dark energy (around 70\%) such as cosmological constant
or quint essence (in agreement with high redshift type Ia supernovae
measurements~\cite{perlmutter,schmidt}). The 30\% of matter consists
mainly of dark matter and the amount of baryons is in agreement with
Big-Bang nucleosynthesis predictions and light elements abundances
measurements. One of the great news coming from these results is that
they agree very well with other measurements of the cosmological
parameters obtained with completely different methods: large scale
structure observation, lensing, type Ia supernovae and light elements
abundances. We are now heading towards a concordance model.

\subsection{Motivations for an intermediate scale experiment}
There is a part of the CMB power spectrum that lacks measurements:
between the large angular scales measurements from COBE (around
$\ell=20$) and the largest angular scales from BOOMERanG, MAXIMA and
DASI (above $\ell\simeq 75$). This can be easily understood as when
you are trying to measure large $\ell$, you concentrate on a small
patch of the sky in order to reach high signal to noise ratio. This
implies loosing all information on large scales. On the other hand,
COBE covered the entire celestial sphere but with a poor 7 degrees
resolution that limited the measurements to the low $\ell$. The
intermediate part of the power spectrum, despite being difficult to
measure, is of high interest.: first, this is the part of the power
spectrum that is the most sensitive to the early Universe physics.
Second, linking COBE measurements to high-$\ell$ measurements with one
single experiment would ensure that no calibration problems are
affecting the data.

Archeops is a balloon-borne experiment designed to measure the CMB
temperature angular power spectrum from the large multi-degree scales
($\ell\simeq 30$) to the small sub-degree scales ($\ell\simeq 800$).
This is achieved via a large sky coverage (around 35\%) along with a
high angular resolution (10 arcminutes). Archeops is also very similar
to Planck High Frequency Instrument and can therefore be considered as
a real size testbed for Planck-HFI.

\section{Archeops instrument}
The Archeops instrumental setup\footnote{see {\tt
    http://www.archeops.org}.}  is exhaustively described
in~\cite{arch_tech} and we will only outline its main characteristics.

\subsection{Telescope and scanning strategy}
\begin{figure}[htb]
\resizebox{\hsize}{!}{\includegraphics{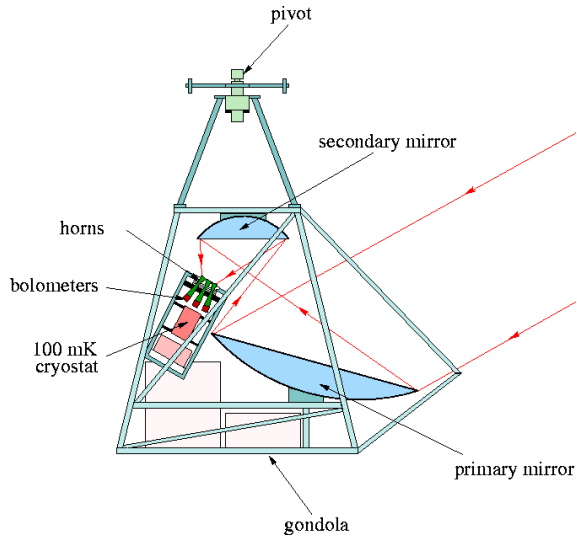}}
\caption{Schematic view of the Archeops gondola. The telescope observes at 
  41 degrees of elevation and rotates at 2 rounds per minute.}
\label{fig_gondola}
\end{figure}

\begin{figure}[htb]
\resizebox{\hsize}{!}{\includegraphics{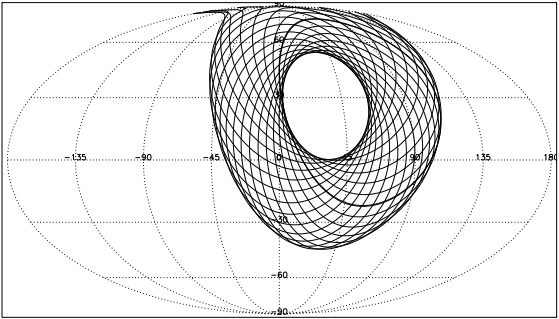}}
\caption{Scanning strategy on the sky (Molleweide projection): a circle 
  is plotted every hour.  After 24 hours, the covered region has the
  shape of an annulus.}
\label{fig_scan}
\end{figure}

The Archeops telescope is an off-axis Gregorian telescope with a 1.5
meter primary mirror that looks at 41 degrees of elevation. During the
flight, the gondola rotates at 2 rounds per minute around its vertical
axis and therefore performs scans on the sky at constant elevation.
While the Earth rotates, these scans end up forming a wide annulus on
the sky with scans crossing each other at various time scales allowing
an efficient monitoring of systematic signals. The gondola is shown in
Figure~\ref{fig_gondola} and the scanning strategy in
Figure~\ref{fig_scan}. The gondola is hanged below a stratospheric
balloon that floats at an altitude of around 40 km. The data is taken
during 24 hours nighttime winter arctic flights.

\subsection{Cryostat, bolometers and cold optics}
The cryostat is a $^3\mathrm{He}-^4\mathrm{He}$ dilution refrigerator
that cools down to 100 mK. This dilution is the same that will be used
for Planck-HFI and does not rely on gravity. The
$^3\mathrm{He}-^4\mathrm{He}$ mixing is made through capillaries all
around the 100 mK stage. The radiation enters through a large (16 cm)
polypropylene window closed by a valve that opens automatically at low
pressure. The radiation is the focused through Planck-HFI-like flared
corrugated back-to-back horns at the 10 K stage. The lower part of the
horns contains the filters in the 1.6 K stage. They filter 4
different frequencies depending on the bolometer: 143, 217, 353 and
545 GHz. The bolometers are located on the focal plane cooled down to
100 mK at the exit of flared horns. The bolometers are spider web
bolometers similar to those that will be used for
Planck-HFI~\cite{mauskopf}. Their nominal noise equivalent power (NEP)
is $1.4\times 10^{-17}\mathrm{W/\sqrt{Hz}}$.

\section{Archeops flights}
The first Archeops test flight took place on July, 18$^\mathrm{th}$
1999 from the Italian balloon launching facility in Trapani (Sicily)
to the South of Spain where the gondola landed on the next day.  This
test flight gave us 4 hours of good quality night-time data with the 6
bolometers that were mounted in the focal plane. This flight helped us
in improving the instrument in order to reduce systematic effects.

The first scientific flight took place in January, 29$^{\mathrm{th}}$
2001 from the SSC\footnote{Swedish Space Corporation.} base of Esrange
(used by CNES\footnote{Centre National d'Etudes Spatiales, the French
  space agency.}) in Kiruna in the North of Sweden. The focal plane
contained 23 bolometers (one of them was blind in order to monitor
systematic effects) with the following repartition: 8 bolometers at
143 GHz, 6 at 217 GHz, 6 at 353 GHz (sensitive to polarization using
orthomode transducers) and 2 at 545 GHz. The interest in covering
various frequencies is that we can monitor this way systematic effects
and astrophysical foregrounds. During the flight, the temperature of
the focal plane remained well below 100 mK (minimum of 89 mK) showing
a perfect behavior of our cryostat. Unfortunately, due to unusually
high stratospheric winds, we used a rather small balloon (150 000 m$^3$
instead of 400 000 m$^3$) and the maximum altitude of the gondola was
31.5 km inducing a higher background and atmospheric contamination.
Also due to these winds, the gondola arrived too fast to our Eastern
limit (Ural mountains) and therefore we had to stop the flight after 7.5
hours at ceiling. This reduced considerably the amount of data taken
during the flight and especially the highly redundant scans that start
after 7 hours of flight. The quality of the data was however excellent
and allows us to perform a CMB anisotropy analysis, especially on the
low $\ell$ edge of the first acoustic peak.

\section{Data analysis}
The data obtained in such an experiment consists of a set of
time streams that represent the value of some detector (thermometer,
bolometer, gyroscope, ...) as a function of time sampled at 153 Hz.
The data analysis is performed through successive steps described in
the following subsection.

\subsection{Pointing reconstruction}
During the flight, the gondola pendulates and does not rotate at
constant speed, we therefore have to reconstruct off line the direction
pointed by each bolometer for each sample. This is done using a
stellar sensor telescope that is attached to the gondola and points in
the same direction as the bolometers. The stellar sensor focal plane
contains 48 photodiodes whose signal exhibits a peak whenever they
cross a star. We therefore obtained an observed star catalog that is
compared with a real one in order to give the pointing solution. Our
pointing accuracy is estimated to be around one arcminute.

\subsection{Data cleaning}
The data coming from the bolometers is characterized by long drifts
that are correlated to the temperature variations inside the cryostat,
pendulations and altitude variations. There is also a scan synchronous
parasitic signal that can be monitored using high frequency channels.
These extra signal are decorrelated and we are left with a time stream
that still exhibits low frequency drifts due to imperfections in the
decorrelation and residual signals.

\subsection{Beam reconstruction}
\begin{figure}[htb]
\resizebox{\hsize}{!}{\includegraphics{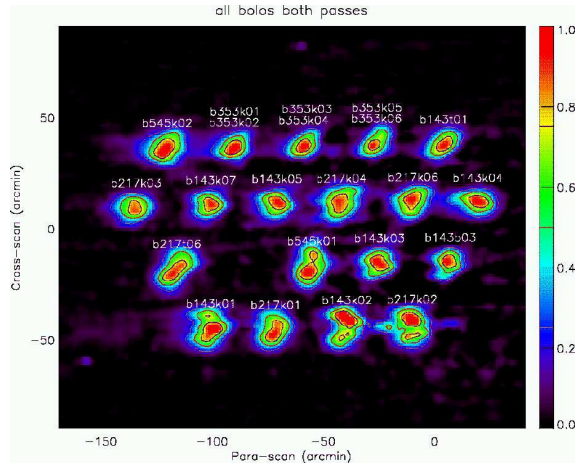}}
\caption{Measured beam shapes for the Archeops bolometers.}
\label{fig_beams}
\end{figure}

During the scientific flight, Jupiter was observed twice with an
interval of roughly two hours. We use Jupiter to obtain maps of our
beams as its angular size (around 40 arcsec) is much smaller than our
angular resolution. The beam maps for all off our bolometers are shown
in Figure~\ref{fig_beams}. We obtain an average angular resolution of
around 12 arcminutes, larger than expected due to the large time
response of the bolometers that can be seen in the figure on left side
of each beam.

\subsection{Calibration}
Jupiter can be used to calibrate our data as its surface brightness
temperature is known to be $170\mathrm{K_{RJ}}$ with an uncertainty of
15\%. This gives us a point source calibration. But the best
calibration is obtained on extended sources like the Galactic plane
and the cosmological dipole that both fill the beam, therefore
including possible sidelobes. We obtained at the end a calibration for
each bolometer with an accuracy of 15\%. The uncertainty is mainly due
to the residual scan synchronous parasitic signal.

\subsection{Power spectrum estimation}
\begin{figure}[htb]
\resizebox{\hsize}{!}{\includegraphics{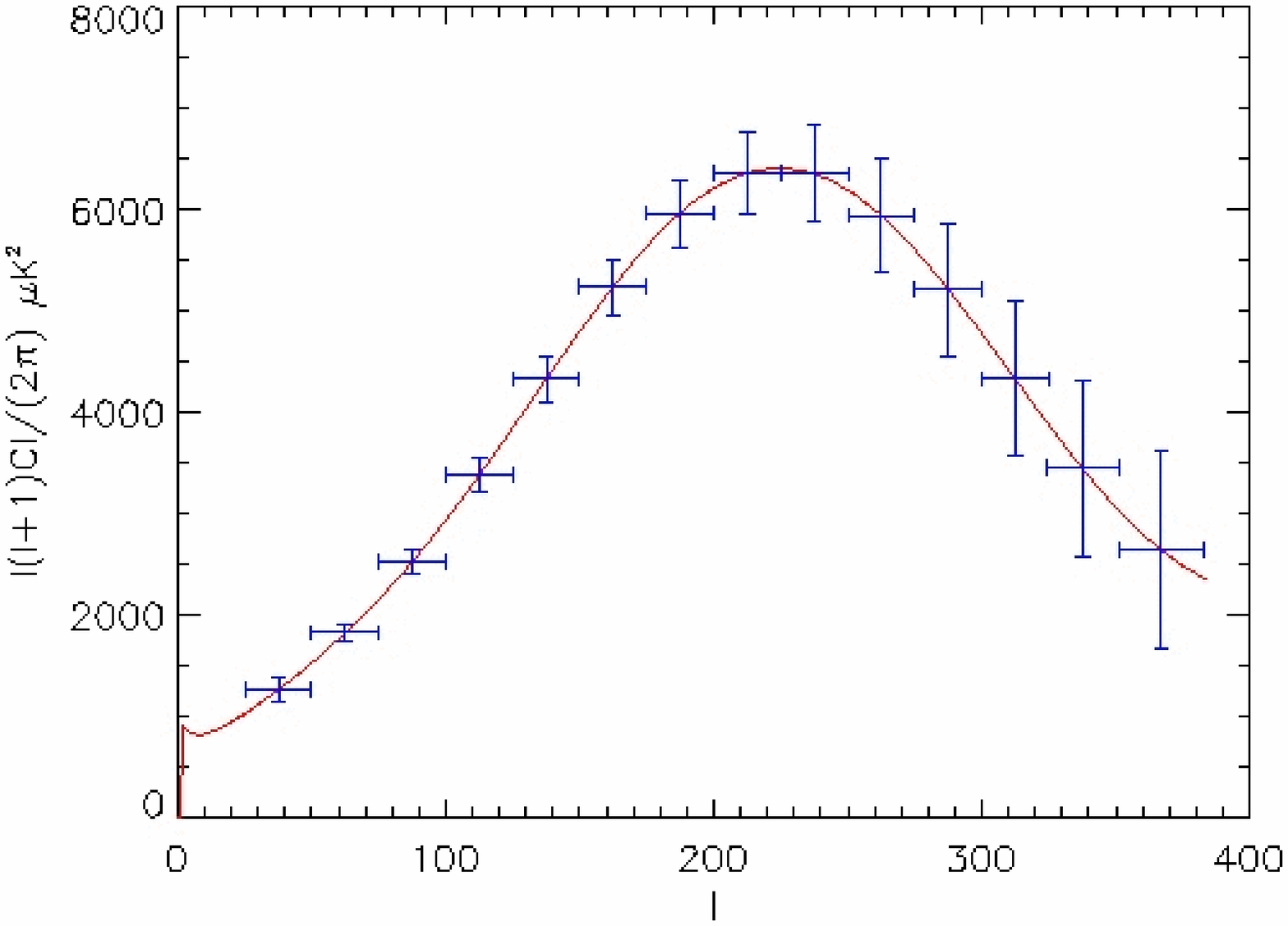}}
\resizebox{\hsize}{!}{\includegraphics{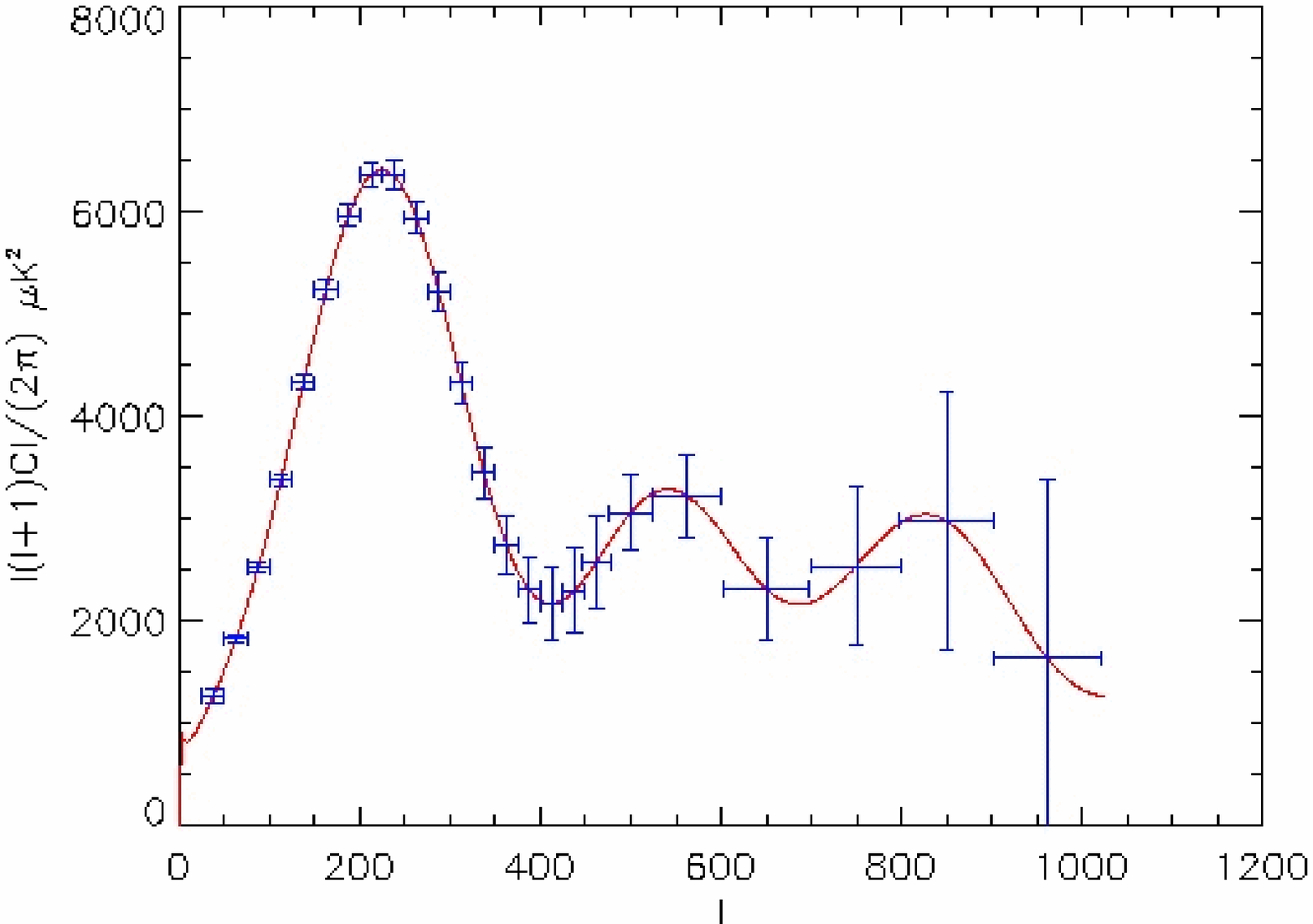}}
\caption{Estimated power spectrum accuracy for the 7h30 Kiruna scientific 
  flight (top) and for the incoming 24 hours flights (bottom). Both
  were obtained for 10 bolometers from the average of one thousand
  realistic simulations of the Archeops timelines and noise
  structure.}
\label{figsim}
\end{figure}

We use the MASTER method~\cite{hivon} in order to measure the power
spectrum on our maps. Within this framework, maps are obtained by
coadding the filtered timelines on the sky. In the case of pure white
noise, this leads to optimal maps. We filter our timelines keeping
only frequencies between 1 and 45 Hz so that the resulting power
spectrum is very close to be white. The pseudo-$C_\ell$ spectrum is
obtained using {\em anafast} in Healpix package~\cite{gorski}. We then
estimate the noise angular power spectrum on the coadded maps using a
set of Monte-Carlo simulations. The effect of the filter on the
underlying sky is also estimated via Monte-Carlo simulations. The mode
mixing effect is deconvolved following~\cite{hivon}. All these
corrections allow us to transform the pseudo-$C_\ell$ spectrum into a
real angular power spectrum. We estimated the non optimality of our
power spectrum (due to the non optimal maps) to be less than 30\% at
all scales. The power spectrum estimation is at the moment under test.
The expected accuracy for 10 bolometers, obtained through full simulations from
timelines to power spectrum with realistic noise (measured on the real
timelines), is shown in top panel of Figure~\ref{figsim}.

\section{Perspectives for incoming flights}
A flight campaign is planned for Archeops this winter. We plan to
flight once in December 2001 and once in January 2002. We will be
able to achieve two 24 hours flights at high altitude (around 40 km).
This will give us much more redundancies than on the previous 7h30
flight and cleaner data due to a higher altitude. The expected
accuracy on the CMB angular power spectrum was estimated using the
noise measured for the last flight, 10 bolometers, but for a 24 hours
flight. The result is shown in the bottom panel of
Figure~\ref{figsim}.

These simulations show clearly that Archeops will be able to measure
precisely the angular power spectrum of the CMB on both small and
large angular scales. The improvement compared  to
previous experiment will be particularly significant on the low-$\ell$
part of the power spectrum (smaller error bars and better
sampling).

\end{document}